\documentclass{PoS}
\usepackage{url}

\title{Unidentified Fermi Objects in the view of H.E.S.S. - Possible Dark Matter Clumps}

\ShortTitle{Unidentified Fermi Objects in the view of H.E.S.S.}

\author{Dorit Glawion$^a$, Denys Malyshev$^b$, \speaker{Emmanuel Moulin}$^c$, Louise Oakes$^d$,  Lucia Rinchiuso$^c$, Aion Viana$^e$, for the H.E.S.S. Collaboration\footnote{for collaboration list see PoS(ICRC2019)1177}   \\
\llap{$^a$}  Landessternwarte, ZAH, Universit\"at Heidelberg\\
D-69117 Heidelberg, Germany\\
\llap{$^b$} Institut f\"ur Astronomie and Astrophysik, Universit\"at T\"ubingen \\
D-72076 T\"ubingen, Germany\\
\llap{$^c$}  IRFU, CEA, Universite Paris-Saclay\\
F-91191 Gif-sur-Yvette, France\\
\llap{$^d$} Institut f\"ur Physik, Humboldt-Universit\"at zu Berlin\\
D-12489 Berlin, Germany\\
\llap{$^e$} Instituto de F\'isica de S\~ao Carlos, Universidade de S\~ao Paulo \\
BRA-13566-590, S\~ao Carlos, Brasil\\

E-mail: \email{dglawion@lsw.uni-heidelberg.de}, \email{dmalishev@gmail.com}, \email{emmanuel.moulin@cea.fr}, \email{loakes@physik.hu-berlin.de}, \email{lucia.rinchiuso@cea.fr}, \email{aion.viana@ifsc.usp.br}}

\abstract{There is strong evidence about the existence of unknown dark matter in the Universe. Many different theories about this dark matter exist, but most probably it is made of a new kind of fundamental particle that has to be massive, stable, electrically neutral, and having only weak interaction with standard matter (weakly interacting massive particles). In principle, those particles could produce gamma rays by their annihilation or decay. Therefore, a Gamma-ray signal from a dark matter origin would provide one of the clearest and most concluding evidences for dark matter. High resolution cosmological N-body simulations have shown that dark matter subhalos in the Milky Way halo may developed in the Universe. Those subhalos could pop-up in gamma-ray surveys as unidentified sources.

In this paper we present H.E.S.S. observations of unidentified sources selected from Fermi-LAT catalogs. These sources fulfill main features which would characterize a dark matter subhalo, namely, having no obvious counterpart at other wavelengths and being steady hard sources}

\FullConference{36th International Cosmic Ray Conference -ICRC2019-\\
		July 24th - August 1st, 2019\\
		Madison, WI, U.S.A.}

\begin{document}

\section{Indirect Dark Matter Search with Gamma-ray Observations}

There is strong evidence about the existence of unknown dark matter in the Universe. Many different theories about this dark matter exist, but most probably it is made of a new kind of fundamental particle that has to be massive, stable, electrically neutral, and undergoes only weak interactions with the standard baryonic matter, and therefore, are called weakly interacting massive particles (WIMPs). Within the $\Lambda$CDM paradigm, at least six times more dark matter than baryonic matter exists in our Universe. Today, understanding this dark matter is by no doubt one of the most important topics of physics \cite{Bertone2005a}. In the search for dark matter three different approaches emerge: the direct production in collider experiments \cite{Kane2008}, the direct detection through scattering off ordinary matter \cite{Cerdeno2010}, and the indirect detection based on the search for secondary particles produced by the annihilation or decay of dark matter particles \cite{Bertone2005b}. There are many theories offering dark matter particle candidates which could annihilate into $\gamma$-ray photons. One of the best is the supersymmetric extension of the Standard Model of particle physics (SUSY) \cite{Wess1974,Haber1985} which provides a natural dark matter particle candidate, the lightest neutralino $\chi$.

A $\gamma$-ray signal from dark matter would provide one of the clearest evidence for dark matter. Spectral features such as annihilation lines \cite{Bertone2009a} and internal Bremsstrahlung \cite{Bringmann2008} as well as a characteristic cut-off at the dark matter particle mass would characterize a dark matter origin, shedding light over the nature of the dark matter constituent. This spectrum must be universal. Hence, a forceful smoking-gun for dark matter would be the detection of several $\gamma$-ray sources, with no counterpart at other wavelengths, all of them sharing identical spectra \cite{Pieri2009,Lee2009,Ando2009,Baltz2000}.

Astrophysical regions where a high dark matter density is foreseen are the best candidates to expect $\gamma$-ray emission from dark matter annihilation or decay. Very high energy $\gamma$-ray ($>100\,$GeV) emission from dark matter annihilation in center of our galaxy is predicted but several bright very high energy sources in the vicinity of the Galactic Center are present that can over-shine the dark matter signal \cite{Aharonian2006,hessDM4}. Proven efficient alternative searches focuses in regions without detected (known) TeV emission \cite{Aharonian2016}. Dwarf spheroidal galaxies are known to have a high dark matter mass and are nearby. Several of these objects have been observed so far in the very high energy regime, but no hint of a signal has ever been found \cite{Aharonian2008, ahnen2008, aleksic2014, Strigaro2018}. Very high energy $\gamma$-ray emission of dark matter origin might be detectable in galaxy clusters despite being very far objects. Unfortunately, the signal may be over-shined by very high energy emission from, e.g., active galactic nuclei. Observations of galaxy clusters have been performed and no dark matter signal was found so far \cite{Aharonian2012, Acciari2018}.

There exist other possible regions of high dark matter density. In the last years cosmological N-body simulations have successfully uncovered how the cold dark matter distribution evolves from almost homogeneous initial conditions into the present hierarchical and highly clustered state \cite{Diemand2008,Springel2008}. High resolution simulations indicate that dark matter halos should not be smooth but must exhibit a wealth of substructures on all resolved mass scales \cite{Kuhlen2008,Stadel2008}. These subhalos could be too small to have attracted enough baryonic matter to start star-formation and would therefore be invisible to past and present astronomical observations. Overdensities or clumps are foreseen into these subhalos which can be nearby, e.g., inside the Galactic halo and therefore bright at very high energies \cite{Pieri2008}. Also dark matter high density regions can develop around intermediate massive black holes from where a rather peaked very high energy emission is predicted \cite{Aharonian2008b,Bertone2009b}. These overdensities would most probably only be visible at high and very high energy gamma-ray band. Because dark matter emission is expected to be constant, such hypothetical sources would pop-up in the all-sky $gamma$-ray programs \cite{Kamionkowski} as unidentified objects, e.g., observed with the \textit{Fermi} satellite and not detected at any other wavelengths.

As already mentioned, the smoking-gun for dark matter detection can be a very distinct energy cut-off close to the dark matter particle mass. Most probably, this is too high in energy \cite{Amsler2010} to be measurable by \textit{Fermi}-LAT within a reasonable time. Therefore, the synergy between \textit{Fermi} and ground based Cherenkov telescopes is mandatory. Furthermore, due to a much larger effective collection area of Cherenkov telescopes, studies of flux variability of the gamma-ray emission of short time scales are more meaningful. In this contribution we present the observations of four unidentified \textit{Fermi} objects from the 3FHL catalog \cite{3FHL} observed with the H.E.S.S. telescopes and discuss the implications for Dark Matter research.

\section{Selection of unidentified \textit{Fermi} Objects}

In order to obtain the best candidates for H.E.S.S. observations, we have performed a thorough selection of steady, hard sources, having no obvious counterpart at other wavelengths in the \textit{Third Catalog of Hard Fermi-LAT Sources} \cite{3FHL}, looking for dark matter clump candidates. The following criteria given in Table~1 were applied. 

\begin{table}[h!]
\small
\label{table:1}     
\begin{center}
\begin{tabular}{l r}     
\hline
Criteria  & No. of sources \\\hline
Without association &  178 \\
Far enough from the galactic plane, cut in galactic latitude of $|b|>5^\circ$  & 126\\
Non-variable, cut in variability 
index (No. of Bayesian blocks in var. analysis) equal to 1  & 125\\
Maximum culmination angle at H.E.S.S. site of $45^\circ$ & 83\\
Follow a simple power law with significance for curvature $<3\sigma$    & 83\\
Hard spectrum, cut in spectral index below 2 & 18\\
No MWL counterparts & 6\\
\hline
\end{tabular}
\end{center}
\caption{Selection criteria applied to 3FHL catalog. For the multi-wavelength (MWL) counterpart search, individual search radii were used ($\sim2-4$ arcmin) based on uncertainty of Fermi position. The following list of MWL facilities were checked: \textit{XMM-Newton}, \textit{ROSAT}, \textit{SUZAKU}, \textit{CGRO}, \textit{Chandra}, \textit{Swift}, WMAP, \textit{RXTE}, \textit{Nustar}, SDSS, \textit{Planck}, \textit{WISE}, \textit{HST}.}
\end{table}

Out of 178 unassociated objects, six objects were surviving these criteria. A list of the these unidentified \textit{Fermi} objects is given in Table~2. The first four in Table~2 were selected for observations with the H.E.S.S. telescopes.

\begin{table}[h!]
\small
\label{table:2}     
\begin{center}
\begin{tabular}{l c c c c c}     
\hline
Source  & Coord. & Pos. unc. & $E_\mathrm{pivot}$ & Diff. Flux at pivot & Pow. law  \\
3FHL..	& RA [h] DEC [$^\circ$] & [arcmin] & [GeV] & [cm$^{-2}$\,GeV$^{-1}$\,s$^{-1}$]&  index    \\
\hline
J1915.2$-$1323	&19:15:16.4 -13:23:30	&3.042	&61.0	&(0.8$\pm$0.3)$\times10^{-13}$	&1.48$\pm$0.33		\\
J0929.2$-$4110	&09:29:17.9 -41:10:10	&2.676	&43.0	&(1.4$\pm$0.6)$\times10^{-13}$	&1.66$\pm$0.37		\\
J2030.2$-$5037	&20:30:16.8 -50:37:50	&2.934	&37.6	&(1.9$\pm$0.8)$\times10^{-13}$	&1.74$\pm$0.33		\\
J2104.5$+$2117	&21:04:34.1 21:17:01	&1.944	&35.1	&(2.6$\pm$1.0)$\times10^{-13}$	&1.80$\pm$0.33		\\
J1553.8$-$2425	&15:53:50.6 -24:25:14	&3.972	&32.4	&(3.5$\pm$1.2)$\times10^{-13}$	&1.85$\pm$0.33		\\
J0813.7$-$0353	&08:13:46.5 -03:53:57	&3.402	&29.4	&(3.5$\pm$1.4)$\times10^{-13}$	&1.93$\pm$0.39		\\
\hline
\end{tabular}
\end{center}
\caption{List of selected candidates with spectral properties given in the 3FHL catalog.  }
\end{table}

\section{H.E.S.S. observations and data analysis}

Observations of unidentified \textit{Fermi} objects have been performed in the very-high-energy ($E>100$\,GeV) gamma-ray range with the High Energy Stereoscopic System (H.E.S.S.) which is an array of five Imaging Athmospheric Cherenkov Telescopes located in the Khomas Highland in Namibia \cite{Aharonian06}. These measurements were conducted in 2018 and 2019 including the four 12\,m telescopes with a mirror area of 108\,m$^2$ as well as with the fifth telescope (CT5) with a mirror area of 614\,m$^2$. 

For all unidentified \textit{Fermi} objects, the telescopes pointed towards the sky direction indicated in the 3FHL catalog in wobble mode \cite{berge2006} with a offset of 0.7$^\circ$.  

The analysis of the data was performed using a Hillas reconstruction technique \cite{parsons2014} and the background consisting of cosmic-ray events is being rejected with a neural network based scheme \cite{murach2015}. We used the \textit{ring} and \textit{reflected-region} method for the calculation of the maps and the differential upper limits, respectively, for the estimation of the residual background contamination level of the source region (number of ON and OFF events) \cite{berge2006}. The value $\alpha_\mathrm{Exp}$ gives the ratio of the on-source time to the off-source time. We analyzed 
data from CT5 together with the smaller telescopes in order to achieve the best sensitivity over a broader energy range and assumed a point-like emission. A cross-check analysis and check of the same data were performed using an additional independent calibration and analysis software \cite{deNaurois09} providing compatible results.

\section{Results}

\begin{table}[h!]
\small
\label{table:3}     
\begin{center}
\begin{tabular}{l c c c c c}     
\hline
Source  & $t_\mathrm{eff}$   & ON  & OFF  & $\alpha_\mathrm{Exp}$  &  Sig.  \\ 
region	& [h]  &  &  &  & [$\sigma$]      \\
\hline
3FHL\,J0929.2$-$4110	&7.8   & 243   & 5458  & 0.046 &  -0.6        \\
3FHL\,J1915.2$-$1323	& 3.0	&95	&2479	&0.045	&-1.5		\\  
3FHL\,J2030.2$-$5037	&8.8	&229	&5325	&0.047	&-1.2		\\
3FHL\,J2104.5$+$2117	&5.5	&102	&2445	&0.044	&-0.6		\\
\hline
\end{tabular}
\end{center}
\caption{Preliminary analysis results of H.E.S.S. data from unidentified \textit{Fermi} objects. Numbers are given for the \textit{ring} background rejection method.}
\end{table}

\begin{figure}
   \centering
   \includegraphics[width=7.cm]{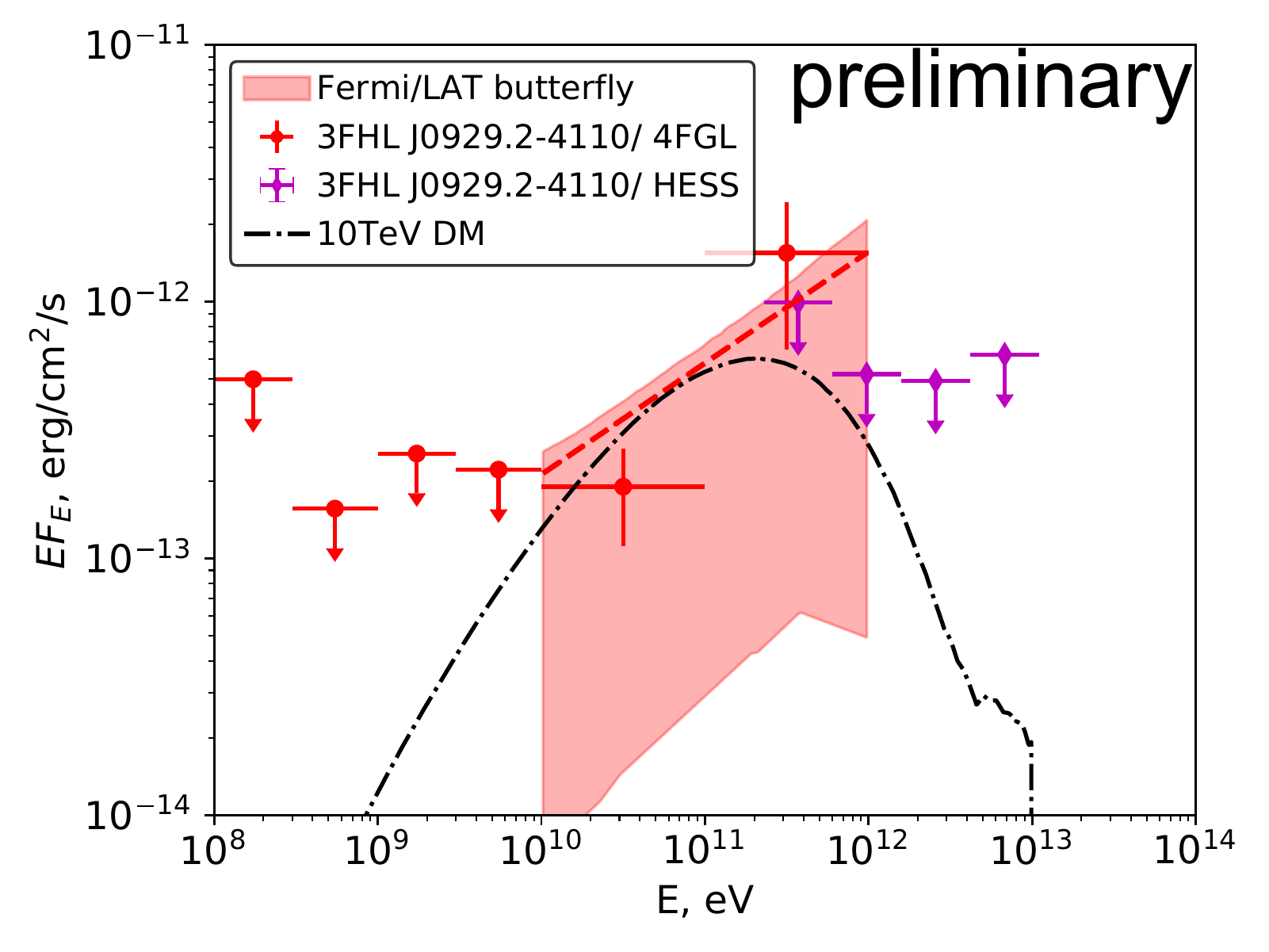}
   \includegraphics[width=7.5cm]{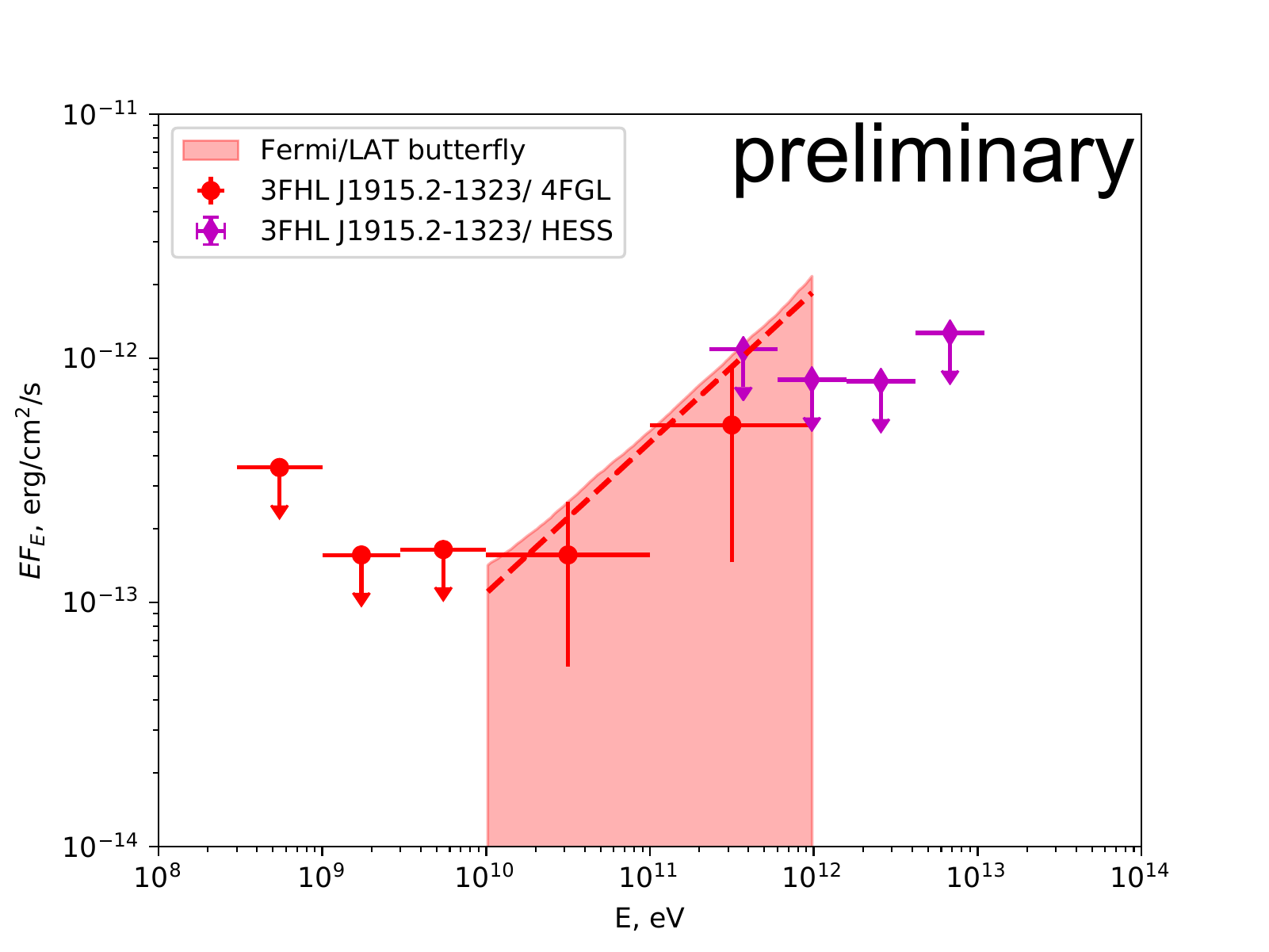}
   \includegraphics[width=7.5cm]{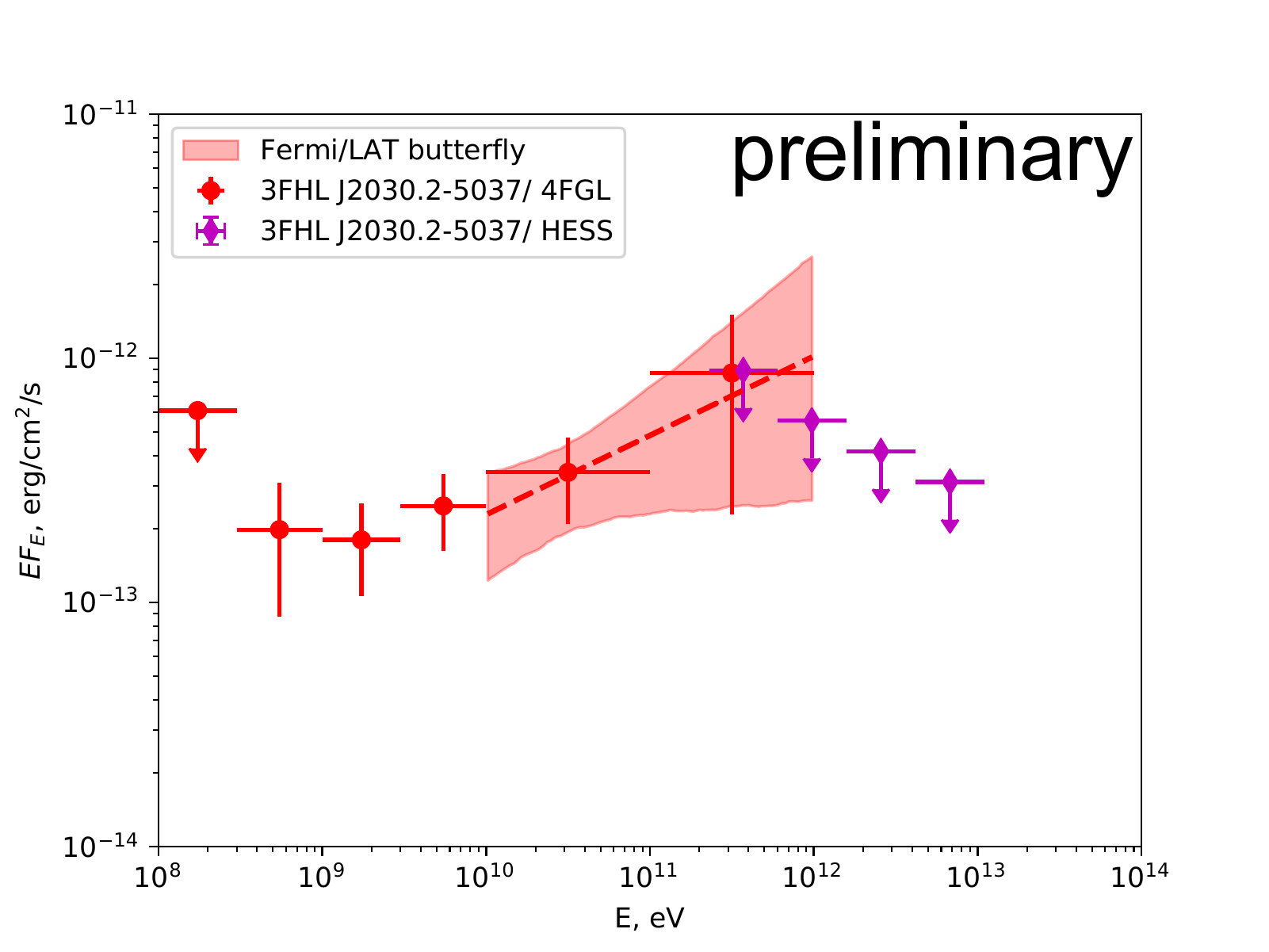}
   \includegraphics[width=7.5cm]{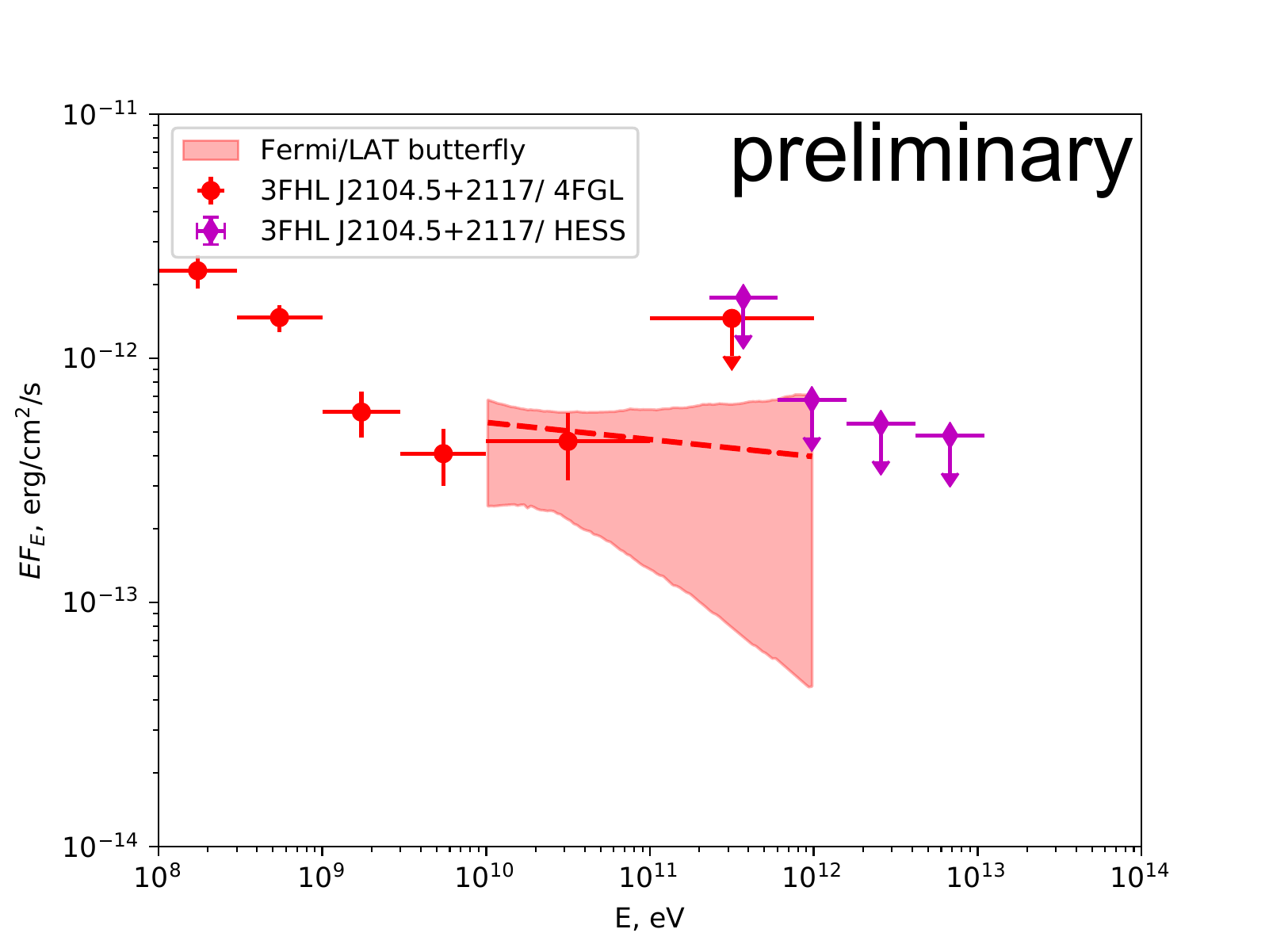}
   \caption{Preliminary spectral energy distributions of different unidentified \textit{Fermi} objects observed with \textit{Fermi}-LAT and H.E.S.S.. Red data points, upper limits, and butterflies show the results from the 4FGL \cite{4FGL} catalog. Magenta 95\% confidence level upper limits give the results from the H.E.S.S. observations.  
   }
              \label{Fig:SEDs}%
    \end{figure}

The resulting numbers for ON and OFF events measured during an effective time $t_\mathrm{eff}$, and the corresponding $\alpha_\mathrm{Exp}$ and significance values for each unidentified \textit{Fermi} object are given in Table~\ref{table:2}. For non of the selected regions a significant point-like emission in the direction of the \textit{Fermi}-LAT positions was found. Therefore, we calculated differential energy upper limits with 95\% confidence level and assuming a spectral photon index of 2.5 and show them in Fig.~\ref{Fig:SEDs} together with the spectral energy distributions for all individual regions as obtained from the 4FGL catalog.

\section{Discussion and Conclusion}

In the following, we discuss the possibility of the dark matter interpretation of detected unidentified \textit{Fermi} objects. The tightest constraints on GeV-TeV mass scale dark matter annihilation cross-section were put \textit{Fermi}-LAT \cite{1507.03530} and the H.E.S.S. collaborations \cite{Aharonian2016} basing on observations of dwarf spheroidal galaxies and the Galactic Center.

Assuming that unidentified \textit{Fermi} objects indeed originate from dark matter annihilation in the dark matter Milky Way clumps, the GeV-TeV spectrum of the signal allows to put constraints on the J-factors of clumps. The GeV-TeV spectral energy distributions of the unidentified \textit{Fermi} objects measured by \textit{Fermi}-LAT and H.E.S.S. provide a characteristic flux level expected from these sources. Combined with existing limits on $<\sigma v>$, these constraints can be converted to constraints on J-factor of the clump.

\begin{figure}
  \centering
  \includegraphics[width=0.7\columnwidth]{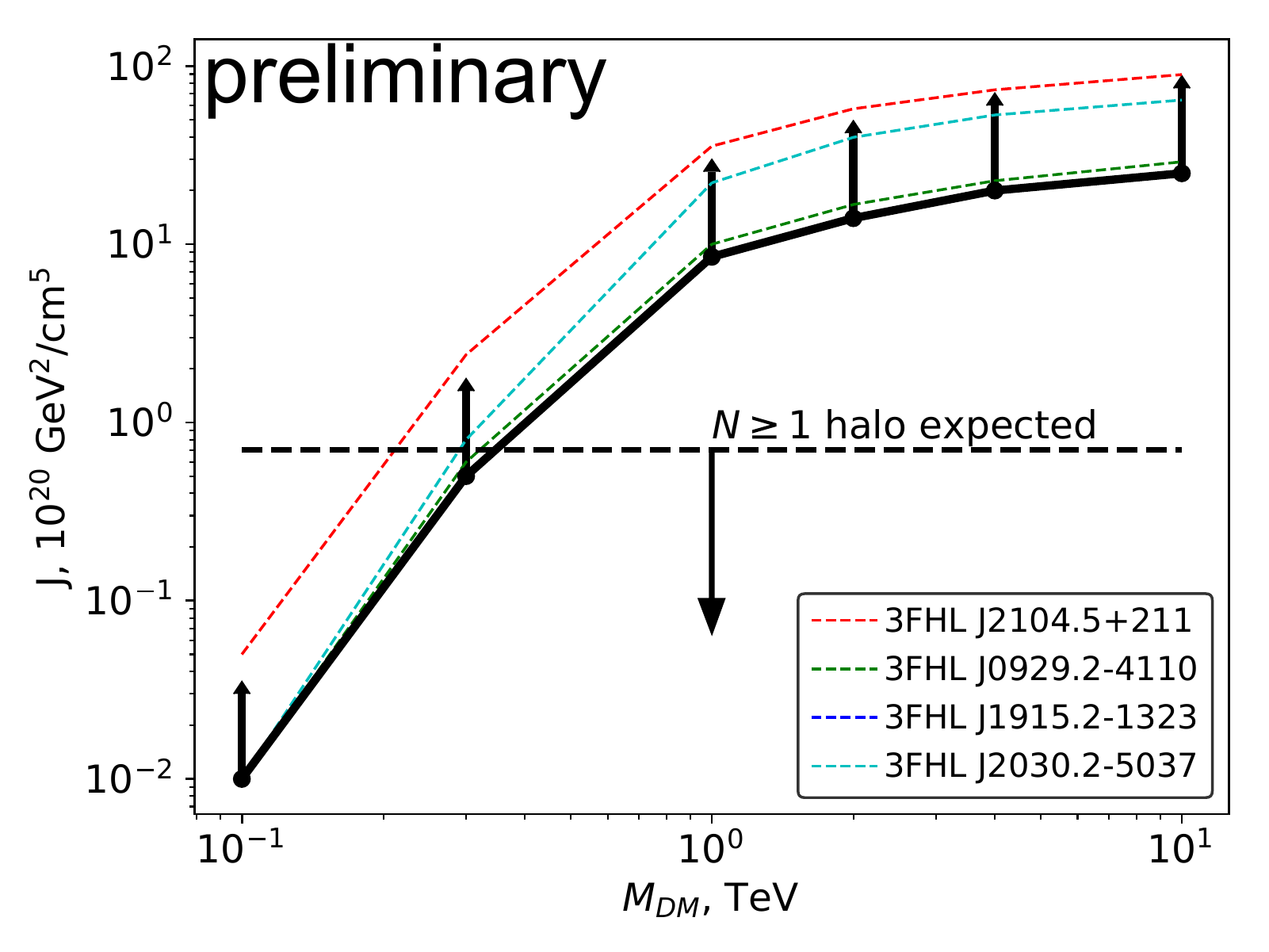}
  \caption{Dark matter clumps interpretation of UFOs allow to put lower limits on $J$-factors of the clumps. Colored dashed lines present result for individual UFOs objects, while solid black line illustrates minimal among UFOs $J$-factor. Horizontal dashed line show $J_{max}$ -- the maximal $J$-factor for which $N\geq 1$ dark matter clumps is present in the Milky Way according to numerical simulations~\cite{1606.04898}. See text for further details. }
  \label{fig:jlimits}
\end{figure}

To illustrate this we consider constraints on annihilation in the $\tau^-\tau^+$ channel. The spectrum of annihilating dark matter with masses in a range of 0.1--100\,TeV was calculated using \texttt{DMFitFunction}, \cite{0808.2641} built in into the \textit{Fermi}-LAT analysis tools. The normalization of the spectrum is proportional to $<\sigma v>\cdot J$ and was selected to fit the \textit{Fermi}-LAT and H.E.S.S. spectral points. A typical spectrum for $M_{DM}=10$~TeV is shown in Fig.~\ref{Fig:SEDs} for the case of  3FHL J0929.2$-$4110.  Existing constraints on $<\sigma v>$ reported in \cite{1507.03530, Aharonian2016} allow to interpret obtained results as \textit{lower limits} on the $J$-factor, see Fig.~\ref{fig:jlimits}. The thin dashed lines show the result for individual unidentified \textit{Fermi} objects, while the thick black line present the lowest $J$-factor value seen at least in one unidentified \textit{Fermi} object. To be on a conservative side in what follow we utilize this line as a lower limit on the DM-clump $J$-factor.

On the other hand, the distribution of clumps in the Milky Way via their J-factors is known from numerical simulations, see e.g. \cite{1606.04898}. The simulations result in a cutoff power law-like distribution of clumps with a strong suppression at most at $J_{max}\sim 0.7\times10^{20}$~GeV$^2$cm$^{-5}$, shown with a dashed black horizontal line in Fig.~\ref{fig:jlimits}. Only $N\ll 1$ clumps with $J>J_{max}$ can be present in the Milky Way.
Comparing the here presented lower limits on $J$-factors to $J_{max}$ we conclude that unidentified \textit{Fermi} objects can be interpreted as clumps of a dark matter only if $M_{DM}\lesssim 0.4$~TeV if we assume $<\sigma v>$ from \cite{1507.03530}.

To summarize our results, we present joint Fermi/LAT and HESS observations of four Unidentified Fermi Objects (3FHL\,J2104.5$+$211, 3FHL\,J0929.2$-$4110, 3FHL\,J1915.2$-$1323 and 3FHL J2030.2$-$5037) which resulted in a detection in GeV and upper limits in TeV band. Combining obtained spectral energy distributions with existing limits on dark matter annihilation cross-section and DM-clumps distribution via their $J$-factors we illustrated that UFOs can be clumps of dark matter only for relatively light dark matter particles with masses $M_{DM}\lesssim 0.4$~TeV.

\section{ACKNOWLEDGEMENTS}

H.E.S.S. gratefully acknowledges financial support from the agencies and organizations listed at \url{https://www.mpi-hd.mpg.de/hfm/HESS/pages/publications/auxiliary/HESS-Acknowledgements-2019.html}.

\end{document}